\documentclass[useAMS,usenatbib]{mn2e}
\usepackage{graphicx}
\usepackage{amsmath}
\usepackage{array}
\usepackage{color}
\usepackage{times}
\usepackage{epstopdf}
\begin{document}

\date{\today}
\pagerange{\pageref{firstpage}--\pageref{lastpage}} 

\label{firstpage}

\title[On the surface extraction of electrons in a pulsar]{On the surface extraction of electrons in a pulsar}%
\author[D. A. Diver et al.]{D.~A.~Diver \thanks{diver@astro.gla.ac.uk for enquiries}, A.~A.~da~Costa$^1$, E.~W.~Laing, C.~R.~Stark$^2$ and L.~F.~A.~Teodoro$^3$\\Department of Physics and Astronomy,
  Kelvin Building,
  University of Glasgow,
  Glasgow, G12 8QQ, Scotland, UK \\
$^1$Sec\c{c}\~{a}o de Telecomunica\c{c}\~{o}es, DEEC, Instituto Superior T\'{e}cnico-UTL, 1049-001 Lisboa, Portugal\\
$^2$School of Mathematics and Statistics, University of St Andrews, Mathematical Institute, St Andrews, KY16 9SS, Scotland\\
$^3$Now at ELORET Corp., Space Sciences and Astrobiology Division, MS 254-3, NASA Ames Research Center, Moffett Field, CA 94935-1000, USA}
\maketitle
\begin{abstract}

We present a novel description of how energetic electrons may be ejected from the pulsar interior into the atmosphere,
based on the collective electrostatic oscillations of interior electrons confined to move parallel to the magnetic
field. The size of the interior magnetic field influences the interior plasma frequency, via the associated matter
density compression. The plasma oscillations occur close to the  regions of maximum magnetic field curvature, that is,
close to the magnetic poles where the majority of magnetic flux emerges. Given that these oscillations have a
density-dependent maximum amplitude before wave-breaking occurs, such waves can eject energetic electrons using only
the self-field of the electron population in the interior. Moreover, photons emitted by electrons in the bulk of the
oscillation can escape along the field lines by virtue of the lower opacity there (and the fact that they are emitted
predominantly in this direction), leading to features in the spectra of pulsars.


\end{abstract}

\section{Introduction and context}
The source of the plasma in pulsar magnetospheres is a subject that has long been at the heart of pulsar electrodynamics, and is a problem still open to question (for example, see review articles by \citet{michel82,michel04}). It is assumed that there must be an initial electron flux from the stellar surface, and that these particles are then implicated in the production of the electron-positron plasma which populates the magnetosphere \citep{1971PhRvL..27.1306R,sokolov,2006RPPh...69.2631H}. The stellar surface in question is the transition between the pulsar interior and its atmosphere; the formation of an environmental pair plasma then must be  linked to the energetics of electrons expelled from the outer crust into the atmosphere.

Therefore we are interested primarily in the physical mechanism that extracts electrons from the interior, and projects them into the atmosphere immediately above the pulsar surface, where the electron-positron plasma is formed. This breaks down the fundamental pulsar problem into two parts:
\begin{enumerate}
  \item From where do the energetic electrons come?
  \item Given these energetic electrons, what underlying physical processes cause the creation of the atmospheric pair-plasma?
\end{enumerate}


This article attempts to contribute a new perspective on (i), the source and energetics of ejected electrons. The central idea is that electrons are ejected from the magnetic flux tubes as a result of the self-field induced in plasma oscillations inside the tubes themselves, by virtue of the magnetic curvature. Electrons are assumed to be severely constrained to follow magnetic field lines in the interior; these lines are strongly curved near the poles, where the magnetic field emerges normal to the pulsar surface (this is because the magnetic field to some extent must run parallel to the curved surface in the interior, but then emerge in a small spot that constitutes the polar region; all emergent field lines are sharply refracted as they leave the iron crustal interior - a shell of iron surrounding a denser neutron core - and emerge into the atmosphere). The curvature of the field along which electrons are constrained to move causes density variations in the electron population, as particles are forced to lose parallel momentum. Such density variations cause electrostatic compressions and rarefactions of the electron `gas' at the plasma frequency, with the self-electric field providing the restoring force (that is, the mutual repulsion of the negatively charged electrons).

These electrostatic waves also have a characteristic radiation signature at the plasma frequency, since the electrons are strongly accelerated when participating in this bulk motion. Given that regions of high magnetic flux have a relatively low opacity for photon transport, this means that photons radiated in this way are more likely to move parallel to the field and emerge at the pole. Given that the plasma density in the pulsar atmosphere is substantially less than that prevailing in the interior, the atmosphere is transparent to these photons, and we present tentative evidence that such signatures are already apparent in the x-ray spectra of selected objects.

The plasma frequency plays a key role here in the electrodynamics, and we show how the plasma frequency must be dependent on the magnetic field, since the latter causes the matter deformation and compression that enables the perfectly conducting, anisotropic transport of electrons along flux tubes to exist.

The sections that follow address key issues in sequence: the magnetic compression, and its influence over the plasma frequency; the dynamics of 1-D electrostatic waves and oscillations; the modelling of the Fermi energy, and the associated maximum energy gain in an electron wave; and a selection of possible candidate pulsars showing non-thermal features that correlate with the prediction of radiation at the internal plasma frequency. Concluding remarks finish the paper, and an explanatory Appendix gives more detail on the relativistic calculation of the Fermi energy.
\section{Density compression and the plasma frequency}
The process of extracting electrons from the outer crust, requires the description of the structure of matter there, and this has to be supported with the study of the structure of atoms in strong magnetic fields.

There are various exotic descriptions of atoms in extreme magnetic fields, showing how the basic lattice structure of
the conducting metal is severely distorted, with implications for the conduction electrons. Such models are
controversial, and in some respects, difficult to reconcile (for example, the literature
\citep{1986PhRvA..33.2084N,1987PhRvA..36.4163N, 1992PhRvL..69..749L}  is undecided about the complex nature of the
bonding between iron atoms under such conditions).

However, the more general concept is widely accepted, namely that the magnetic compression of the atoms leads to an
effective iron atomic radius $R$ given by \citep{2001RvMP...73..629L}
\begin{equation}
R\approx Z^{1/5}b^{-2/5}a_0
\end{equation}

where $Z$ is the atomic number (26 for Fe), $b=B/B_0$  is the ratio of the pulsar magnetic field to the critical field
$B_0=m_e^2e^3c/\hbar ^3=2.35\times 10^5$T and $a_0= 5.29 \times 10^{-11}$m is the Bohr radius. For the typical pulsar
field of $10^8$T,  $b\approx 426$ and $R\approx 9\times 10^{-12}$m. The classical radius of Fe is
$R_\text{Fe}=1.4\times 10^{-10}$m \citep{1964JChPh..41.3199S}, so that the compression leads to an increase in matter
density of $(R_{Fe}/R)^3\approx 3.75\times 10^3$. Of course, the free electron density is also increased as a direct
consequence of this compression. Taking the free electron density of Fe under standard terrestrial conditions as
$n_e^\text{Fe}=1.7\times 10^{29}$m$^{-3}$ \citep{ash+merm}, we can arrive at an electron number density for the pulsar
$n_{e}^{A}$ from atomic compression by applying the same scaling:
\begin{equation}
n_{e}^{A} \approx  n_e^{\text{Fe}}\times 3.75\times 10^3\approx 6.4\times 10^{32}\mbox{m}^{-3} \label{ne_atom}
\end{equation}
which is a lower limit, since it is likely that there are more electrons able to access the conduction band under compression than in the classical limit.

However, there is another way of deriving the empirical electron number density in the pulsar crust. Iron has a monatomic body-centred cubic (bcc) crystal structure under standard terrestrial conditions, with inter-atomic spacing $a_{\text{Fe}}=2.87\times 10^{-10}$m \citep{ash+merm}. Assuming that the iron on the surface of the pulsar is also bcc \citep{1971PhRvL..27.1306R}, this means that the compression factor here can be calculated in terms of the effective atom spacing in the pulsar environment compared with the terrestrial one.

The presence of extraordinarily large magnetic fields distorts the iron crystal from being isotropic to being severely anisotropic in transport terms: the iron is very highly conducting in the direction parallel to the magnetic field, but perpendicular transport is severely inhibited. The overall picture is of a set of perfectly conducting `tubes'  aligned locally with the ambient magnetic field direction along which electrons are able to move relatively freely; this description is qualitatively consistent with the `distorted atoms in a strong field' model. This is essentially the structure proposed by \citet{1971PhRvL..27.1306R};  \citet{1977FCPh....2..203C} show that the conductivity parallel to the magnetic field is on average 20 times greater than that of the field-free case; conversely, the transverse conductivity has a strong dependence on the magnetic field and the Fermi energy, and is typically orders of magnitude smaller than the longitudinal case.


With this simple picture we can capture the essence of the electron motion; parallel momentum is unconstrained, but perpendicular momentum is quantised. In Ruderman's simple one-dimensional flux tube \citep{1971PhRvL..27.1306R}, the radius of the flux tube is set equal to the mean orbital radius associated with the Landau ground state:
\begin{equation}
\hat{\rho}=\left(\frac{\hbar}{eB}\right)^{1/2}\approx 2.6 \times 10^{-12}\,\mbox{m} \label{rudermanrho}
\end{equation}
where we have taken $B=10^8$T.

In so doing, the simple 1-D flux tube model assumes dominant parallel motion along the magnetic field, and assumes that the transport anisotropy is sufficient to render all non-trivial Landau levels as unimportant, effectively confining the electron to a single flux tube.

The scale-length $\hat\rho$ is then the effective inter-atom spacing in the compressed bcc structure, leading to a compression of $(a^\text{Fe}/\hat\rho)^3\approx 1.4\times 10^6$, and an electron number density $n_{eL}$ based on lattice compression of
\begin{equation}
n_{e}^{L}\approx n_e^\text{Fe}\times 1.4\times 10^6\approx 2.4\times 10^{35}\,\mbox{m}^{-3}\label{ne_lattice}
\end{equation}
Since we now have possible electron number densities ranging over more than 2 orders of magnitude, we shall take the geometric mean as the characteristic pulsar interior electron number density $n_e$:
\begin{eqnarray}
n_e^A &=&n_e^\text{Fe}\chi_A b^{6/5}\approx 6.4\times 10^{32}\,\mbox{m}^{-3}\\[4pt]
\chi_A&=&R_\text{Fe}^3 Z^{-3/5}a_0^{-3}\approx 2.61\nonumber\\[4pt]
n_e^L &=&n_e^\text{Fe}\chi_L b^{3/2}\approx 2.4\times 10^{35}\,\mbox{m}^{-3}\\[4pt]
\chi_L &=&a_\text{Fe}^3(eB_0/\hbar)^{3/2}\approx 159\\[4pt]
n_e &=&(n_{e}^{A}n_{e}^{L})^{1/2}=n_e^\text{Fe}(\chi_A\chi_L)^{1/2}b^{1.35}\nonumber\\[4pt]
&\approx& 3.5\times 10^{30}\times b^{1.35}\,\mbox{m}^{-3}\nonumber\\[4pt]
&\approx&1.23 \times 10^{34}\,\mbox{m}^{-3}\label{ne_geom}
\end{eqnarray}
where we have assumed a magnetic field of $B=10^8$T, that is, $b\approx 426$.

The geometric mean compression yields a mass density of approximately $10^6$kgm$^{-3}$ for $^{56}$Fe, which is consistent with the conventional assumptions of density on the surface \citep{shapiro,2006RPPh...69.2631H}.

The electron number density is a critical parameter for plasmas, since it defines the basic collective timescale, namely the plasma frequency $\omega_p$. Given that the interior is an excellent example of fixed positive ions and mobile electrons, we can define the plasma frequency here in terms of the electrons only:
\begin{eqnarray}
\omega_p &=& \left[\frac{n_e e^2}{\epsilon_0 m_e}\right]^{1/2}\\
&\approx&  1.05 \times 10^{17}b^{0.675}\label{plasfreq}
\end{eqnarray}
where we have used the geometric mean number density in the numerical evaluation; $m_e$ and $\epsilon_0$ are the electron rest mass and electric constant, respectively.

\textcolor{black}{Notice that the plasma frequency in the pulsar interior depends on the magnetic field strength; it is not an
independent parameter. This is an important result that, although appearing to be a straightforward deduction from the mass density dependence well-known in the literature, is still worth stressing because of the radiative consequences.}

\textcolor{black}{Having used a blend of the published equations of state for the bound material in the lattice to infer the free electron density, it is important to recognise that such free electrons are not required to obey the same equation of state as the bound matter: after all, the electrons in a cold plasma have no equation of state! Since the cold electron plasma is an excellent model for free electron behaviour in our pulsar interior, we shall therefore not require any further detailed analysis of the interior equation of state.}

\section{1-D electron dynamics}
\subsection{Calculating the Fermi energy}
The simple assumption of such a strongly anisotropic conducting flux tube has major implications. The electron dynamics in the interior can be treated essentially as one-dimensional, since the electron momentum parallel to the magnetic field greatly exceeds that perpendicular to it; hence in modelling the electron distributions, and deriving the Fermi energy as a substitute for the surface work function, a one-dimensional treatment will be a good approximation. Moreover, since cross-field transport is inhibited, the role of the Landau levels in the interior is diminished: in the non-relativistic formulation of the electron motion in a uniform strong magnetic field $B$ \citep{sokolov}, the electron energy $\varepsilon$ is given by
\begin{equation}
\varepsilon\approx m_ec^2+p_{||}^2/(2m_e) + (n+\tfrac{1}{2})\hbar\omega_c
\end{equation}
where $\omega_c=eB/m_e$ is the cyclotron frequency, $m_e$ is the electron rest mass and $n=0,1,2,...$ is the principal quantum number for the quantised electron orbit in the plane perpendicular to the magnetic field. The radius $R$ of orbit of the perpendicular motion of the electron can be expressed (in the same non-relativistic limit) as \citet{sokolov}
\begin{equation}
R=\left[\frac{2(n+\tfrac{1}{2})\hbar}{eB}\right]^{1/2}
\end{equation}

The mean orbital radius associated with the Landau ground state is given by
\begin{equation}
R(n=0)=\left(\frac{\hbar}{eB}\right)^{1/2}\approx 2.6 \times 10^{-12}\,\mbox{m}
\end{equation}
in other words, $\hat{\rho}=R(n=0)$, in Eq.~(\ref{rudermanrho}), as already assumed.
Note that the energy increment $\delta \varepsilon$ between Landau levels is
\begin{equation}
\delta \varepsilon = \frac{\hbar e}{m_e}B\approx 10^{-4}B\quad\mbox{eV} \label{ll_deltaE}
\end{equation}
where $B$ is given in Tesla. Hence for a typical pulsar magnetic field of $10^8$T, Landau levels are separated by  $\sim 11.5$keV, implying that there is negligible population of the higher Landau levels if thermal excitation is the only mechanism, given that the typical surface temperature of a pulsar is $<10^6$K (in energetic terms, $< 100$eV) \citep{2007Ap&SS.308..181H,kargaltsev2007,2006ApJ...646.1104B}. This reinforces the merit in assuming that the electron momentum is largely parallel to the magnetic field, and lends credence to the argument that a one-dimensional statistical treatment captures the essential physics.

Finally, although calculations involving the Landau levels are quantum in nature, they are not relativistic (though the generalisation is possible). For the moment, we will defer the detailed discussion about the need for a relativistic treatment to the Appendix.

In general, the distribution function for electrons in the presence of a magnetic field $B$  is really the Fermi-Dirac distribution, $f_{FD}$,
\begin{equation}
f_{FD}=\left\{\exp[(\epsilon\pm \mu_B B- \mu)/(k_BT)]+1\right\}^{-1}
\end{equation}
where $\epsilon$ is the energy, $\mu$ is the chemical potential, $\mu_B$ is the magnetic moment of the electron , and $T$ is the temperature. In the limit $T\rightarrow 0$, $\mu \rightarrow \epsilon_F$, the Fermi energy,  and since $\mu_B B\sim 6\, {\mbox keV}\gg k_BT$ for $B\sim 10^8$T, we can assume that the electrons are spin-aligned in the lowest energy configuration, and so this term can be neglected. Hence we have
\begin{eqnarray}\label{M06}
    \lim_{T\rightarrow 0}f_{FD}&\simeq&\lim_{T\rightarrow 0}\left\{\exp[(\epsilon - \mu)/(k_BT)]+1\right\}^{-1}\\
    &=&\left\{\begin{array}{ll}
                        1&\mbox{for $\varepsilon<\mu$}\\
                        0&\mbox{for $\varepsilon>\mu$}
                  \end{array}\right. \nonumber
\end{eqnarray}
For temperatures such that $k_BT \ll \mu$, the distribution is therefore basically a step function, with all energy levels equally occupied up to $\mu$, and none of the higher ones occupied.

 \citet{1971PhRvL..27.1306R} calculated the Fermi energy for the simple 1-D case, motivated by the
restricted motion of the electrons imposed by the enormous magnetic field strengths in the pulsar interior. The reason for calculating the Fermi energy is that $\epsilon_F$ is an excellent guide to the work function of the surface, that is, the potential barrier which must be surmounted before interior particles can escape to the exterior.

Assuming that the mean electron energy is far below the Fermi temperature (so that the step-function nature of the Fermi distribution can be assumed), for $N$ electrons in the population,
\begin{equation}
N=\int_0^\infty g(\epsilon)f_{FD}(\epsilon)\mbox{d}\epsilon \approx \int_0^\mu g(\epsilon)\mbox{d}\epsilon \label{genfermi}
\end{equation}
where $g(\epsilon)$ is the density of states for the electron gas, and $\mu = \epsilon_F$ is the Fermi energy in the limit of $T \ll T_F$.
For the one-dimensional electron gas, $g(k)\mbox{d}k= [L/(2\pi)]\mbox{d}k$, where $L$ is the characteristic scale-length of the problem, and we have suppressed the normal degeneracy factor $2$, given the spin-alignment assumption. In this case, the integration yields
\begin{equation}
\epsilon_F = \frac{h^2N^2}{2m_eL^2}
\end{equation}
where we should interpret $N/L$ as the line-density of electrons. Assuming a uniform density approximation within the pulsar interior, the line density of electrons confined to a flux-tube of radius $\hat{\rho}$ is simply the volume electron number density $n_e$ times the cross-sectional area associated with the flux-tube: $N/L = n_e \pi \hat{\rho}^2$. When substituted, this gives
\begin{eqnarray}
\epsilon_F=\frac{h^2n_e^2\pi^2\hat{\rho}^4}{2m_e}&=&\frac{h^4}{8m_ee^2}\frac{n_e^2}{B^2}\label{fermi1dprim}\\
&\approx& 1.03\times 10^{-66} \frac{n_e^2}{B^2}\nonumber\\
&\approx&2.28\times 10^{-16}\,b^{0.7}\,\mbox{J}\label{efermi1dclassical}
\end{eqnarray}
in SI units. Eq.~(\ref{fermi1dprim}) is essentially Ruderman's formula \citep{1971PhRvL..27.1306R}.

Assuming a typical magnetic field of $10^8$T, and taking the electron number density from (\ref{ne_geom}), yields
\begin{equation}
\epsilon_F \approx 97\,\mbox{k\,eV}\label{ef_simple}
\end{equation}

Notice that the Fermi energy increases with magnetic field strength: this is because the electron number density increases faster than the field strength, no matter which scaling model is chosen. This is an important, if slightly counter-intuitive, result, since the influence of the magnetic field in expressions for $\epsilon_F$ such as Eq.~(\ref{fermi1dprim}) suggest that an increasing magnetic field strength will lower the potential barrier at the poles. However, this neglects the indirect influence of $B$ on the electron number density: increasing field strengths must lead to increased matter compression, and this must be taken into account in the Fermi energy calculation.

 Note that since the surface temperature of the pulsar $T_p$ is below $10^6$K, equivalent in energy terms to $<100$eV, then (\ref{ef_simple}) is an acceptable approximation, since in this case it is self-evident that $T_p \ll T_F$.

 The calculations presented here assume a 1-D model for electron motion, and therefore preclude any contribution from Landau levels. The Appendix details the nature of the calculation where Landau levels must be included, and where the energies are intrinsically relativistic.

\subsection{Electron motion in inhomogeneous magnetic fields}

The simple 1-D description of the electron transport requires careful treatment when the magnetic field is  not homogeneous. Magnetic field emerges from the interior predominantly at the poles; this concentration of the field lines at specific points on the surface means that there must be significant curvature inside (and outside) the star, as the field emerges from inside the iron spherical shell. Electrons constrained in the 1-D model to follow these field lines are nevertheless inertial particles, and therefore cannot instantaneously change direction. However, they are also restricted in their perpendicular motion, since they have to satisfy the Landau quantization. In the simple 1-D model, the electrons negotiate this magnetic field curvature in a non-quantal way, since scattering into Landau levels above the ground state is not permitted; therefore the electrons must change direction to follow the field direction, without any significant locally transverse excursions; momentum conservation must be provided by the crystal lattice, and the overall energy conservation must lead to electrons losing energy to the crystal as they negotiate the curve.

The overall effect of the magnetic inhomogeneity must be to introduce a local `bunching' of electron density along the magnetic field direction in the region of greatest local curvature, that is, near the magnetic poles.
\textcolor{black}{Note that complex magnetic geometries can be produced by thermal evolution of the neutron star itself \citep{page, pons2009}, where strong toroidal field curvature can result from the strong magnetic feedback on thermodynamic transport properties. If strongly curved magnetic fields produced in such process lead to significant perpendicular field eruption at the pulsar surface, then these field lines will also act as electron ejection sources.} A schematic of the situation is shown in Figure 1. Such local density fluctuations will drive an electrostatic wave along the flux tube, accelerating (and decelerating) non-local electrons (ahead and behind the curved region) that populate this tube.  Now either this is sufficient in itself to eject electrons at the end of the tube, where the surface exists, or the electrostatic wave itself evolves non-linearly and eventually breaks, leading to the ejection of a few relatively energetic particles ahead of the wave. Either way, it seems that such a description has the requisite element of a parallel acceleration mechanism that is self-consistent, and not dependent on any frame-transformed field component.

\begin{figure}
\resizebox{\hsize}{!}{\includegraphics{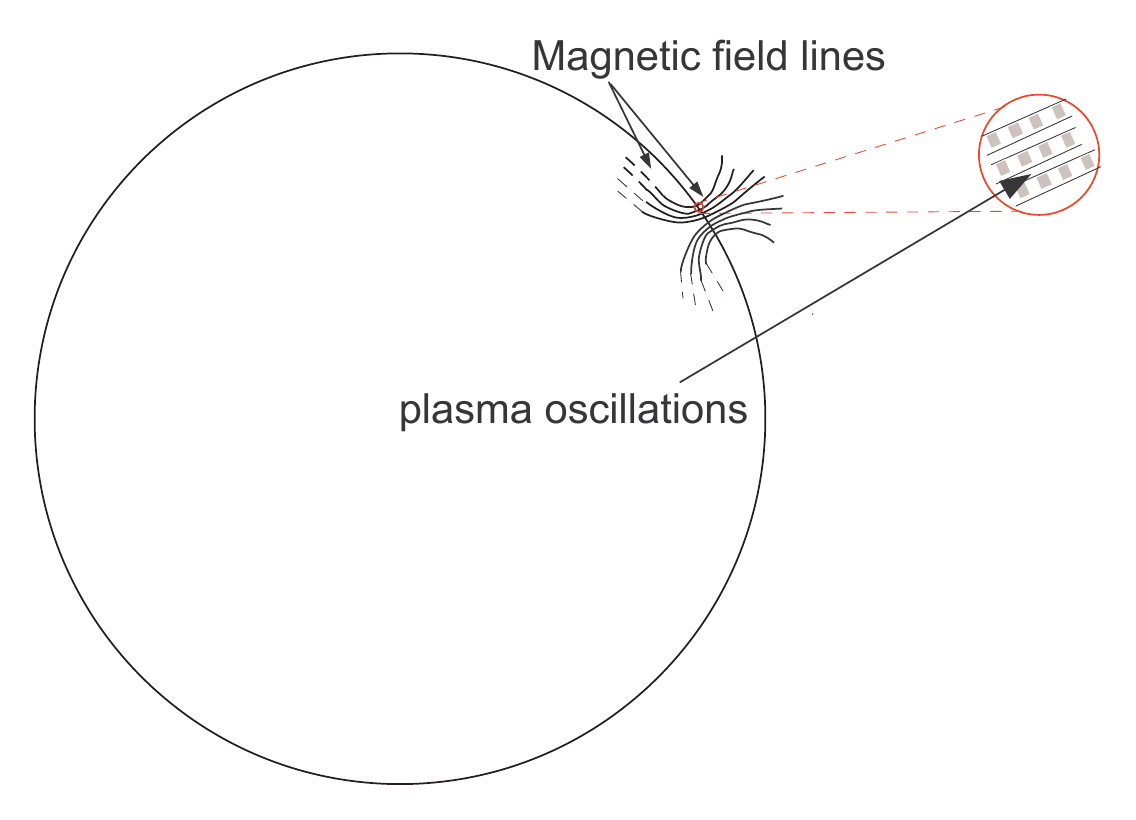}} \caption{Schematic showing a simplified sketch of part of the magnetic field of the pulsar, interior and exterior (note that only a few magnetic field lines are included, in order to keep the diagram simple). A small region near the pole is magnified to show the location of the electron oscillations induced by the curvature; the shaded regions are intended to illustrate density compressions. \textcolor{black}{Note that strong magnetic field curvature can also be caused by magneto-convective processes not represented in this diagram.} }
\end{figure}

It is worth distinguishing between the electron temperature and the electron energy; mono-energetic beams of electrons are formally cold (that is, possess zero temperature), yet can stream in the beam direction with significant energy. In the scenario outlined above, it is assumed that the mean energy of the electrons, including any directed streaming, is less than the Fermi energy, since otherwise a significant fraction of the interior electrons would simply leak out of the flux tube into the pulsar atmosphere at the surface. Whilst this might solve the electron production problem, there remains the issue of identifying the physical process that causes such streaming. For the purposes of this article, therefore, we will assume that the electrons are mostly trapped in the iron crust, just as the free electrons are trapped in a metal in a natural terrestrial context. As in the latter case, there needs to be a mechanism that provides the necessary energy for electrons to overcome the potential barrier at the interface between the pulsar surface and the vacuum (or atmosphere), before electrons can be liberated at the surface.  Thermionic emission seems implausible, since the surface temperature is less than one tenth of the Fermi temperature; hence our concentration on the (self) electric field produced by charge concentrations of electrons in density waves.

On a more formal basis, electrons with energy less than the escape energy (that is, the Fermi energy) that are directed towards the surface along the curving magnetic field lines must be reflected, since they cannot escape. Therefore there must be two populations of electrons in motion along a flux tube: those that are moving towards the surface, and those that are moving towards the interior, having been reflected at some point. This is similar to the magnetic bottle effect \citep{2003phpl.book.....B} where charged particles are reflected at magnetic field concentrations; note that the analogy is not perfect, though, since in the magnetic bottle, particles are reflected simply because energy is transferred from parallel motion along the field to perpendicular motion in the form of larmor orbits. Such counter-streaming electrons will lead to density instabilities if the perturbations on each stream happen to evolve in phase such that they reinforce each other.
The full dispersion relation for electrons moving parallel and anti-parallel to the magnetic field direction with speed $u$ is given by \citet{2003phpl.book.....B}
\begin{equation}
\frac{\omega_1^2}{(\omega - ku)^2}+\frac{\omega_2^2}{(\omega+ku)^2}=1 \label{twostream}
\end{equation}
in which $\omega_1^2=n_1e^2/(\epsilon_0m_e)$ is the square of the plasma frequency for electrons moving parallel to the
magnetic field with speed $u$, where $n_1$ is the number density of such electrons; $\omega_2^2$ is similarly defined
for electrons moving anti-parallel to the field. We have assumed symmetry in the modelling, for simplicity; clearly we
expect $n_1\approx n_2$, which is equivalent to saying that the electron loss-rate is small; hence we also expect $\omega_i \approx \omega_p$, $i=1,2$.  The frequency and wavenumber of the common perturbation on the electron streams are given by $\omega$ and $k$ respectively. We must solve Eq. (\ref{twostream}) in order to arrive at the criteria for instability.  Manipulation of the algebra reveals that the dispersion relation is biquadratic in $\omega$, and that there are imaginary components of frequency (or wavenumber) when $u<\sqrt{2}\omega_p/k$.  Put another way, any density perturbations with a wavelength $\lambda$ satisfying
\begin{equation}
\frac{2\pi u}{\sqrt{2}\omega_p}<\lambda<R_*
\end{equation}
must be unstable, where we have taken the upper limit on the wavelength to be the maximum scalelength in the star itself, that is, the pulsar radius $R_*$. Since the lower limit is at most $2\pi c /(\sqrt{2}\omega_p)\approx 5 \times 10^{-10}$m $\approx 200\hat{\rho}\ll R_*$, then it is always possible for instabilities to affect the counterstreaming flows, in the form of longitudinal plasma oscillations at the plasma frequency.

There are further assumptions in this simple treatment, namely that the plasma density is uniform, and the electrons
are monoenergetic. These assumptions are fine in the context of an idealised cold plasma treatment; we contend that
they serve as an adequate illustration in the context of the pulsar flux tube.

The nature of this streaming instability is to induce electrostatic oscillations on the dynamic electron population
inside the flux tube.

These plasma oscillations produce a self electric field because of the compression and rarefaction of electron density;
there is no associated magnetic perturbation, since the particle and displacement currents cancel perfectly. Note that
this is a pure electron plasma with a fixed positive ion background, and so density enhancements here are pure electron
overdensities. There is however a maximum amplitude of electric field associated with such oscillations. If the
wavenumber is $k$, then in the non-relativistic limit, this maximum electric field $E_{\text{max}}$
\citep{1990PhST...30....5K,1990PhST...30..127M}:
\begin{equation}
E_{\text{max}} = \frac{m_e\omega_{p}^2}{ek}\label{emaxnonrel}
\end{equation}
This result can be extended to the relativistic case:
\begin{equation}
E_{\text{max}}= \sqrt{2}\frac{\omega_{p
}m_ec}{e}(\gamma_\phi-1)^{1/2}
\end{equation}
where $\gamma_\phi = (1+p_\phi^2/(m^2c^2))^{1/2}$ is the relativistic factor for maximum associated electron momentum $p_\phi$ in the oscillation. This is the limiting field before coherent motion breaks down and wavebreaking occurs, at which point part of the electron population begins free-streaming at speeds greater than the phase velocity of the oscillation itself \citep{1990PhST...30..127M,1990PhST...30....5K}.

Consider the case of an electrostatic oscillation in a flux tube, very close to the surface near the magnetic pole. If the oscillation is at the point of wavebreaking, then the maximum energy $\delta\varepsilon$ that can be gained by an electron falling through the self-field of the oscillation before free-streaming must be given by
\begin{eqnarray}
\delta \varepsilon &\approx& e E_{max} \times \,\mbox{distance travelled by electron}\nonumber\\
&\approx& e E_{max} \pi/k\nonumber
\end{eqnarray}
This electron will escape to the surface if it has an energy greater than the Fermi energy. Hence for escaping electrons,
\begin{equation}
e E_{max} \pi/k>\epsilon_F
\end{equation}
which can be expressed as
\begin{equation}
k^2< \frac{\pi m_e\omega_p^2}{\epsilon_F}\approx 5.6\times 10^{20}b^{0.65}\,\mbox{m}^{-2}
\end{equation}
using Eqs. (\ref{ne_lattice}), (\ref{plasfreq}) and (\ref{efermi1dclassical}). For the typical pulsar magnetic field, $b\approx 426$, yielding $k<1.7\times 10^{11}$m$^{-1}$. In other words, plasma oscillations with wavelengths (or scale variations)$\lambda=2\pi/k$ exceeding $4 \times 10^{-11}$m can eject electrons from the surface, providing such oscillations are at maximum amplitude. This minimum wavelength is considerably greater than $\hat{\rho}\approx 2.6\times 10^{-12}$, the width of the flux-tube (and the effective inter-ion spacing), suggesting that whilst the breakdown of ultra-short wavelength oscillations can't provide sufficient energy for electrons to escape the surface, longer wavelength oscillations will readily do so.

Thus we have shown that instabilities in the electron dynamics within the flux tube are sufficient to eject electrons from the pulsar interior.

Given that the opacity of the pulsar to photons of frequency $\omega$ is reduced by a factor $(\omega/\omega_c)^2$ for propagation parallel to the magnetic field direction \citep{2001RvMP...73..629L,1975NYASA.257..108C}, where $\omega_c=eB/m$ is the electron cyclotron frequency, there is the enhanced prospect of photons characteristic of the oscillation within the flux tube escaping directly to the pulsar atmosphere; in this context, $\omega = \omega_p$, and $(\omega_p/\omega_c)^2\approx 6.4b^{-0.65}\approx 0.12$ for magnetic fields of $10^8$T, leading to the enhanced probability that the magnetic polar regions are transparent to the emission of keV photons (from the plasma oscillation) produced inside the flux tubes by the bulk plasma dynamics. Note that for magnetic fields of around $4\times 10^6$T, this opacity enhancement disappears, and so direct emission of photons below $0.4$keV by this process wouldn't be expected.

\textcolor{black}{There has been considerable discussion of the electromagnetic transmission properties of the pulsar crust, particularly in the strong-field limit relevant to magnetars, in which the free electrons in the crust influence the transmission of the blackbody radiation from the pulsar surface itself (see for example \cite{adelsberg2006, adelsberg, turolla}). Strictly, this article presents a discussion of the electrostatic (not electromagnetic) behaviour of electrons oscillating at the interior plasma frequency. These electrons radiate in the local oscillating electric fields close to the surface such that some of the radiated photons escape from the interior, and are able to propagate in the pulsar environment; a fraction of them may survive atmospheric processes to reach the observer. It is helpful therefore to clarify that the characteristic non-thermal signature of such a process as described in this article is not affected by the cold plasma twin-mode electromagnetic absorption properties of the crust as discussed in the literature.}

Of course, the picture is further complicated by the fact that the self-field of the oscillation can accelerate particles in the perpendicular direction \citep{craigm}. Whilst the 1-D model forbids perpendicular transport within the flux tube inside the pulsar, this restriction is no longer valid at the transition to the atmosphere \textcolor{black}{(since the confining crystal lattice is no longer imposing dynamical constraints)}, and it is possible that some of the emitted electrons will possess a perpendicular velocity component relative to the magnetic field direction very close to the surface, leading to a spread in exit trajectories for escaping electrons.

\section{supporting features in neutron star X-ray spectra}

We offer the following examples from XMM-Newton observations in support of our hypothesis that there is non-thermal
radiation at the internal plasma frequency present in the pulsar emissions. \textcolor{black}{The motivation for searching for such a non-thermal signature in X-ray spectra is that its presence in an object's spectrum lends plausibility to the electron production process actually occuring in that object.}

It is established in the literature that X-ray pulsars commonly have a significant soft X-ray excess \citep{hickox}, but not all such excess signals come from accretion processes (see for example points 3-5 in the conclusions of \citealt{hickox}). We offer here some examples in which we suggest that certain X-ray specral features can be associated with the novel physical processes described in this article. Sifting all available data in support of
non-thermal emission that could be attributed to plasma frequency radiation from the poles is a daunting task, not just
because such signals are magnetic field and geometry dependent, and may in fact lie outside the spectral range for
XMM-Newton. However, we did find several examples, which are described below. The magnetic field strengths in each example (other than in the case of \citet{2004A&A...418..625D}) were taken from the Ioffe pulsar catalogue \footnote{http://www.ioffe.ru/astro1/psr-catalog/Catalog.php}.

Please note that the predicted plasma frequencies are pulsar surface values; terrestrial observation of such features will incur a gravitational redshift of up to 25\%.

De Luca and co-workers \citep{2004A&A...418..625D} showed detailed observations of the x-ray spectrum of the neutron star 1E 1207.4-5209, principally to demonstrate the presence of cyclotron absorption features at 0.7, 1.4, 2.1 and 2.8 keV, from which was derived a surface magnetic field strength of $B\approx 6\times 10^6$T, significantly different from the field of $2.6 \times 10^8$T derived conventionally from timing parameters. The interpretation of the higher harmonics has been challenged \citep{2005ApJ...631.1082M} as an instrumental effect. However, assuming that the lower magnetic field strength is correct, and using the plasma frequency given by Eq.~(\ref{plasfreq}) for this field strength, we calculate $\omega_p\approx 9.4\times 10^{17}$rad s$^{-1}$, generating photons with energy $h\nu_p$ of approximately $0.62$keV. This is very close to an excess above blackbody radiation in the phase integrated emission in the band 0.1-2.5 keV, just before the absorption feature at 0.7keV (see Figure 6 in \citealt{2004A&A...418..625D}). Caution must be exercised here, since there is also a sharp drop in the CCD efficiences in both the pn and MOS1 instruments at 0.5eV \citep{kirsch}; however, since the observations show a rise above the blackbody spectrum before 5eV, followed by a sharp drop, it is reasonable to assume that the excess is genuinely observed.

PSR J0538+2817 \citep{2003ApJ...591..380M, 2004MmSAI..75..458Z} has a surface magnetic field strength of $7.33 \times 10^7$T, giving $h\nu_p \approx 3.3$keV. The energy spectrum of this pulsar shows a marked excess over the blackbody emission at 3\,keV (see Figure 4 in \citealt{2003ApJ...591..380M}) consistent with non-thermal emission from the pole via the mechanism reported here. There are no calibration issues at these energies with the PN and MOS1 instruments on XMM-Newton, both of which recorded the excess signal.

The surface magnetic field of Geminga \citep{2004MmSAI..75..458Z} is $2\times 10^8$T, implying that radiation from the sub-surface plasma oscillation will be at energies $\sim 6.6$keV. The fitted EPIC count-rate spectrum shows a clear departure from the model blackbody spectrum in the range $6-8 \mbox{ keV}$, as shown in Figure 6 of \citet{2004MmSAI..75..458Z}.

PSR J1932+1059 (B1929+10) shows a small feature at 2keV (\citet{2003pasb.conf...37W}, Figure 4) which is not fitted by either the blackbody spectrum or the power law. The pulsar has a surface field of $B_s=5.18\times 10^7$T, suggesting that the plasma frequency yields photons of around 2.6 keV, offering a plausible explanation of this spectral feature.

If the identified spectral features are associated with the plasma frequency, then this means that there are indeed plasma oscillations present at the poles, and therefore it is also possible that such oscillations are contributing to the production of high-energy electrons in the atmosphere. \textcolor{black}{Whilst it is arguable that the features we are citing here are close to being borderline in terms of statistical significance, we contend nevertheless that they offer a tantalising glimpse of what might be the seat of the electron expulsion mechanism, and as such, merit attention.}

\section{Concluding discussion}

The general scenario presented in this paper is as follows: plasma oscillations are induced in the throat of the tightly-curving magnetic field structure as it exits the pulsar interior at the magnetic poles. Such oscillations can sustain maximum amplitude electric fields such that at wavebreaking, electrons can be accelerated to energies in excess of the Fermi energy and ejected from the interior into the immediate pulsar environment. This behaviour could be intermittent, since the transit time for electrons to move along a magnetic field line from one pole to the other is less than a typical pulsar period (less than a ms, in fact), and so there could be an influence on the sub-pulse structure.

In addition to ejecting energetic electrons, the accelerations associated with these interior plasma oscillations will give rise to radiated photons which escape along the low opacity field lines and emerge at the magnetic poles. Since the electron density immediately above the magnetic pole (produced by the ejection of electrons from a lossy mirror) is much less than that in the interior, any plasma oscillation radiation that leaks out will propagate away - photons of around $10^{17}$Hz (around 2 keV) could be directly emitted from the polar region by interior oscillations - another source of moderate energy radiation. We have presented evidence of X-ray spectra of pulsars which show behaviour of this kind.

Given that the self-field in the interior of the flux tube is the source of the acceleration, this field will not be uniformly aligned with the magnetic field, leading to the possibility of highly-energetic imperfectly aligned electrons very close to the pulsar surface. Such electrons could gyro-rotate (subject to Landau levels) and radiate, or perhaps even collide with iron nuclei; either way, energetic photons are possible directly from the surface.

Of course, the pulsar still loses electrons this way; charge balance must be maintained ultimately, and this is
probably achieved by a combination of sucking in electrons from the electron cloud above the poles (each half-cycle the
self-field of the oscillation will reverse) and surface diffusion of electrons back into the interior from the
exterior. Note that any positive potential arising from electron depletion of the pulsar will increase the work
function (as distinct from the Fermi energy), making the potential barrier higher for electrons to escape the surface.
It is conceivable that such charging considerations could be sufficient to suppress electron ejection at the surface,
even to the point of influencing the pair-production in the atmosphere.

This general picture has several favourable aspects: not only does it provide a plausible source of parallel acceleration, leading to electron production at the magnetic poles, but also there is the real possibility of producing azimuthal structure in the electron production at the surface, arising from the mutual interaction of waves in adjacent flux tubes. Moreover, this model is consistent with the ideas associated with the distortion of atoms under strong magnetic fields giving rise to sheaths of free electrons aligned with those fields; it also yields accelerating parallel electric fields, albeit from a different physical process.

It is also worth noting that dropping the magnetic field strength raises the Fermi energy at the surface (since the strong field is suppressing the work function); this in turn reduces the spectrum of electrostatic oscillations that can eject free electrons, reducing the efficiency of atmospheric electron production and the associated energetic beaming processes. This may give an insight into transient pulsar behaviour, for those objects with weaker surface fields

\section{Acknowledgements}
DAD gratefully acknowledges funding from the UK Science and Technology Funding Council (STFC/F002149/I), as does CRS (PP/E001122/1). AAdaC is grateful to the Funda\c{c}\~{a}o para a Ci\^{e}ncia e Tecnologia, Portugal, under Grant SFRH/BSAB/771/2007, who sponsored his sabbatical leave; thanks are due equally to the Instituto Superior T\'{e}cnico and to the Department of Physics and Astronomy, University of Glasgow, for respectively granting and hosting AAdaC's sabbatical leave. LFAT is grateful to the Leverhulme trust for funding at Glasgow University. Finally, thanks are due to H E Potts, C S Maclachlan and E  Bennet for their helpful comments and stimulating discussion.  The Department of Physics and Astronomy in the University of Glasgow is a member of the Scottish Universities Physics Alliance (SUPA). We are also grateful to the anonymous referee, whose valuable feedback helped improve this article.


\begin{thebibliography}{99}

\bibitem[\protect\citeauthoryear{van Adelsberg}{2006}]{adelsberg2006}
{van Adelsberg}~M. and {Lai}~D.,~2006,~
\newblock {MNRAS}, 373,~1495--1522

\bibitem[\protect\citeauthoryear{van Adelsberg}{2005}]{adelsberg}
{van Adelsberg}~M., {Lai}~D., {Potekhin}~A.~Y. and {Arras}~P., 2005,~
\newblock {ApJ}, 628,~902--913

\bibitem[\protect\citeauthoryear{Ashcroft \& Mermin}{1976}]{ash+merm}
{Ashcroft}~N.~W. and  {Mermin}~N.~D.,~1976,
\newblock  {Solid State Physics}. Holt, Rinehart \& Winston, New York
\newblock

\bibitem[\protect\citeauthoryear{Bogdanov et al.}{2006}]{2006ApJ...646.1104B}
{Bogdanov}~S.,  {Grindlay}~J.~E.,   {Heinke}~C.~O.,  {Camilo}~F.,
  {Freire}~P.~C.~C. and {Becker}~W.,~2006,
\newblock {ApJ}, 646,~1104--1115

\bibitem[\protect\citeauthoryear{Boyd \& Sanderson}{2003}]{2003phpl.book.....B}
 {Boyd}~T.~J.~M. and  {Sanderson} J.~J.,~2003,~
\newblock {The Physics of Plasmas}.
\newblock Cambridge University Press, ~UK

\bibitem[\protect\citeauthoryear{Canuto}{1975}]{1975NYASA.257..108C}
{Canuto}~ V.,~1975,
\newblock in ~{Canuto}~V. ed, {Role of Magnetic Fields in Physics and
  Astrophysics}, vol 257, {New York Academy Sciences Annals},
  108--126

\bibitem[\protect\citeauthoryear{Canuto \& Ventura}{1977}]{1977FCPh....2..203C}
{Canuto}~V. and {Ventura}~J., 1977,
\newblock { Fundamentals of Cosmic Physics}, 2,~203--353

\bibitem[\protect\citeauthoryear{De Luca et al.}{2004}]{2004A&A...418..625D}
{De Luca}~A., {Mereghetti}~S., {Caraveo}~P.~A.,  {Moroni}~M., {Mignani}~R.~P.
  and  {Bignami}~G.~F., 2004,
\newblock {A\&A}, 418,~625--637

\bibitem[\protect\citeauthoryear{Geppert}{2004}]{Geppert2004}
{Geppert}~U., ~{K\"{u}ker}~M. and {Page}~D., 2007,~
\newblock {Astrophysics and Space Science}, 308,~181--190

\bibitem[\protect\citeauthoryear{Haberl}{2007}]{2007Ap&SS.308..181H}
{Haberl}~F., 2007,~
\newblock {Astrophysics and Space Science}, ~308,~181--190

\bibitem[\protect\citeauthoryear{Harding \& Lai}{2006}]{2006RPPh...69.2631H}
{Harding}~A.~K. and  {Lai}~D.,~2006,~
\newblock{Rep. Prog. Phys.}, 69,~2631--2708

\bibitem[\protect\citeauthoryear{Hickox et al.}{2004}]{hickox}
{Hickox}~R.~C., {Narayan}~R. and {Kallman}~T.~R.,~2004,~
\newblock{ApJ}, 614,~881--896

\bibitem[\protect\citeauthoryear{Kargaltsev \& Pavlov}{2007}]{kargaltsev2007}
{Kargaltsev}~O. and {Pavlov} ~G.,~2007,~
\newblock {Astrophysics and Space Science}, 308,~287--296

\bibitem[\protect\citeauthoryear{Kirsch}{2004}]{kirsch}
{Kirsch} ~M., 2004,~
XMM-Newton epic status of calibration and data analysis.
\newblock http://xmm.vilspa.esa.es/docs/documents/CAL-TN-0018-2-3.pdf,
  Garching (ESA)

\bibitem[\protect\citeauthoryear{Kruer}{1990}]{1990PhST...30....5K}
{Kruer}~W.~L., 1990,~
\newblock {Physica Scripta Volume T}, 30,~5--9

\bibitem[\protect\citeauthoryear{Lai}{2001}]{2001RvMP...73..629L}
{Lai}~D., 2001,~
\newblock {Rev. Mod. Phys.}, 73,~629

\bibitem[\protect\citeauthoryear{Lieb et al.}{1992}]{1992PhRvL..69..749L}
 {Lieb},~E.~H.,  {Solovej}~J.~P., and {Yngvason}~J., 1992,~
\newblock {Phys. Rev. Letts.}, 69,~749--752

\bibitem[\protect\citeauthoryear{McGowan et al.}{2003}]{2003ApJ...591..380M}
 {McGowan}~K.~E.,  {Kennea},~J.~A., {Zane},~S.,  {C{\'o}rdova},~F.~A., {Cropper},~M.,~
  {Ho},~C.,~ {Sasseen},~T. and  {Vestrand},~W.~T., 2003,~
\newblock {ApJ},~591,~380--387

\bibitem[\protect\citeauthoryear{MacLachlan et al.}{2009}]{craigm}
{MacLachlan}~C.~S.,  {Diver}~D.~A.  and  {Potts}~H.~E., 2009,~
\newblock {New J. Phys}, ~11, ~063001


\bibitem[\protect\citeauthoryear{Michel}{1982}]{michel82}
{Michel}~F.~C., ~1982,~
\newblock{ Rev. Mod. Phys.}, 54,~1--66

\bibitem[\protect\citeauthoryear{Michel}{2004}]{michel04}
{Michel}~F.~C.,~2004,~
\newblock{Advances in Space Research}, 33,~542--551

\bibitem[\protect\citeauthoryear{Mori \& Katsouleas}{1990}]{1990PhST...30..127M}
{Mori}~W.~B.  and {Katsouleas}~T.,~1990,~
\newblock {Physica Scripta Volume T}, 30,~127--133

\bibitem[\protect\citeauthoryear{Mori et al.}{2005}]{2005ApJ...631.1082M}
{Mori}~K., {Chonko}~J.~C. and  {Hailey}~C.~J.,~2005,~
\newblock {ApJ}, 631, 1082--1093


\bibitem[\protect\citeauthoryear{Neuhauser et al.}{1986}]{1986PhRvA..33.2084N}
{Neuhauser}~D., {Langanke}~K.~ and  {Koonin}~S.~E.,~1986,~
\newblock {Phys. Rev. A}, 33,~2084--2086

\bibitem[\protect\citeauthoryear{Neuhauser et al.}{1987}]{1987PhRvA..36.4163N}
{Neuhauser}~D., {Koonin}~S.~E. and {Langanke}~K.,1987,~
\newblock {Phys. Rev. A}, 36,~4163--4175

\bibitem[\protect\citeauthoryear{Page}{2004}]{page}
{Page}~D.,~{Geppert}~U. and {K\"{u}ker}~M.,~2007,~
\newblock {Astrophysics and Space Science}, 308,~403--412

\bibitem[\protect\citeauthoryear{Pons}{2009}]{pons2009}
{Pons}~J.~A., {Miralles}~J.~A.~ and {Geppert}~U., ~2009,~
\newblock {A\& A}, 496,~207--216

\bibitem[\protect\citeauthoryear{Ruderman}{1971}]{1971PhRvL..27.1306R}
{Ruderman}~M.,1971,
\newblock {Phys. Rev. Letts.}, 27,~1306--1308

\bibitem[\protect\citeauthoryear{Shapiro \& Teukolsky}{1983}]{shapiro}
 {Shapiro}~S.~L. and  {Teukolsky}~S.~A.,~1983,~
\newblock {Black Holes, White Dwarfs and Neutron Stars},
\newblock John Wiley \& Sons, Inc., New York

\bibitem[\protect\citeauthoryear{Slater}{1964}]{1964JChPh..41.3199S}
 {Slater}~J.~C.,~1964,~
\newblock {J. Phys. Chem.}, 41,~3199--3204

\bibitem[\protect\citeauthoryear{Sokolov \& Ternov}{1986}]{sokolov}
{Sokolov}~A.~A.  and {Ternov}~I.~M.,~1986,~
\newblock {Radiation from Relativistic Electrons},
\newblock AIP Translation Series. American Institute of Physics, New York

\bibitem[\protect\citeauthoryear{Turolla et al.}{2004}]{turolla}
{Turolla}~R., {Zane}~S. and  {Drake}~J.~J.,~2004,~
\newblock {ApJ} 603,~265--282

\bibitem[\protect\citeauthoryear{Wozna et al.}{2003}]{2003pasb.conf...37W}
{Wozna}~A., {Kuiper}~L.~ and {Hermsen}~W.,~2003,
\newblock in {Cusumano}~G., {Massaro}~E. and {Mineo}~T., eds, {
  Pulsars, AXPs and SGRs Observed with BeppoSAX and Other Observatories, Proc. Intl. Workshop, Marsala, Sept 23-25, 2002, pp.37--44}

\bibitem[\protect\citeauthoryear{Zavlin \& Pavlov}{2004}]{2004MmSAI..75..458Z}
{Zavlin}~V.~E.  and  {Pavlov}~G.~G.,~2004,
\newblock { Memorie della Societa Astronomica Italiana}, 75,~458

\end{thebibliography}

\section{Appendix}
\subsection{Calculation of Fermi Energy in a Landau Levels environment}
The general case should take into account the effect of Landau levels on the Fermi energy. In fact the Fermi energy is dependent on the linear density of the electrons along the magnetic induction field lines, and the regime becomes relativistic above certain values. Following this idea the energy $\epsilon=\gamma mc^2$ is given by
\begin{equation}\label{M01}
    \gamma=\left\{\begin{array}{ll}
                        \sqrt{1+\displaystyle{\frac{p^2_\|}{m^2 c^2}}+2 n\hbar\frac{eB}{m^2 c^2}}&
                        \mbox{spin up}\\
                        \sqrt{1+\displaystyle{\frac{p^2_\|}{m^2 c^2}}+2\left(n+1\right)\hbar\frac{eB}{m^2 c^2}}&
                        \mbox{spin down}
                  \end{array}\right.
\end{equation}
Then we see that the same value of $\gamma$ has two possible values $(n$ and $n+1)$ with the exception $n=0$. Then the total number of electrons is given by
\begin{equation}\label{M03}
    N=\frac{L}{2\pi \hbar}\sum_n\int  \frac{dp_\| }{\exp\left[{\displaystyle\frac{\varepsilon(n,p_\|)-\mu}{kT}}\right]+1}
\end{equation}
because the transverse moment is filled in Landau levels.

The value of the Fermi energy is $\varepsilon_F=\lim_{T\rightarrow 0}\mu$ and therefore (\ref{M06})
 implies, using $p_\|=\varpi_\| mc^2$, that there are upper limits for $n$ and $\varpi_z$. Putting $\varepsilon=\epsilon-mc^2=\varepsilon_1 mc^2$ and $\mu=\mu_1 mc^2,\, \kappa=\hbar\omega/mc^2$ with
\begin{equation}\label{M07}
    \varepsilon_1=\sqrt{1+\varpi^2_\|+n\kappa}-1
\end{equation}

The upper limits of $\varpi_{\|}$ are reached when $\mu_1=\varepsilon_1$
and therefore the limits of integration are

\begin{eqnarray}
  \varpi_\| &=& \left\{\begin{array}{ll}
                        \sqrt{\displaystyle{\mu_1^2+2\mu-2  n   \kappa }}&\mbox{spin up}\\
                        \sqrt{\displaystyle{\mu_1^2+2\mu-2 (n+1)\kappa }} &
                        \mbox{spin down}
                  \end{array}\right. \\
  n &\in&  \left[0,{\mathrm int}(\frac{\mu^2_1+2\mu_1}{\kappa})\right]
\end{eqnarray}

After some algebra we find that
\begin{eqnarray}
   \frac{N}{L} \frac{h}{mc}&=& \sum_0^{\mbox{int}[(\mu^2_1+2\mu_1)/\kappa]}[\sqrt{\mu^2_1+2\mu_1-n\kappa} \\
   &&\quad\qquad+\sqrt{\mu^2_1+2\mu_1-(n+1)\kappa} \label{M12}
\end{eqnarray}
an equation in $\mu_1$ whose solution has to be found numerically, provided
\begin{equation}\label{M121}
   \frac{\mu^2_1+2\mu_1}{\kappa}>1.
\end{equation}

However if only $n=0$ is present, this equation turns into
 \begin{equation}\label{Q12}
    \frac{N}{L} \frac{h}{mc}= \sqrt{\mu^2_1+2\mu_1}+\sqrt{\mu^2_1+2\mu_1-\kappa}
\end{equation}
with general solution
\begin{equation}\label{M14}
    \mu_1=-1+\left[{1+\left(\displaystyle\frac{N}{2L} \frac{h}{mc} +\kappa \frac{L}{2N} \frac{mc}{h}\right)^2}\right]^{1/2}
\end{equation}
If $\displaystyle\left(\frac{N}{2L} \frac{h}{mc} +\kappa \frac{L}{2N} \frac{mc}{h}\right)^2\ll 1$, then Eq.~(\ref{M14}) reduces to
\begin{equation}
\mu_1\approx \displaystyle\frac{1}{2}\left(\frac{N}{2L} \frac{h}{mc} +\kappa \frac{L}{2N} \frac{mc}{h}\right)^2\label{m14a}
\end{equation}
whereas in the limit $\left(\displaystyle\frac{N}{2L} \frac{h}{mc} +\kappa \frac{L}{2N} \frac{mc}{h}\right)^2\gg 1$
\begin{equation}
\mu_1\approx \displaystyle\frac{N}{2L} \frac{h}{mc} +\kappa \frac{L}{2N} \frac{mc}{h}\label{m14b}
\end{equation}

Eq.~(\ref{m14a}) is similar to Ruderman's \citep{1971PhRvL..27.1306R} expression for Fermi energy when we neglect the spin effects and the zero order energy. This means it can only be applied to subrelativistic regimes. However the general solution Eq.~(\ref{M14}) also contains the relativistic case for high density electrons or large magnetic field amplitudes.

The overall solution shows that the  value of the Fermi energy depends on the value of the magnetic field, and if
\begin{equation}\label{M141}
    \left(\frac{N}{2L} \frac{h}{mc}\right)^2 < \frac{\hbar eB}{m^2c^2}
\end{equation}
it may even be the dominant term determining its value. However for the particular case of exclusively longitudinal motion, there should not be any dependence on the magnetic field; that it enters the Fermi energy even in this case is puzzling.

Moreover the combination of a quantum statistical description in the perpendicular plane and a classical one along the magnetic field leads to the situation in which the statistics may preclude occupation of any Landau levels other than the zero one, as given in Eq.~(\ref{M121}).   These contradictions between the description in Landau Levels and the underlying statistics  have to be resolved in future.

\label{lastpage}
\end{document}